\documentclass[preprint,12pt]{elsarticle}

\usepackage{graphicx}

\usepackage{amssymb, amsthm}

\biboptions{compress}

\journal{\hspace{-3.5cm} \raisebox{-1mm}{\begin{tikzpicture} \draw [fill = white, white] (0, 0) rectangle (3.5, 0.4); \end{tikzpicture}}}

\usepackage{amsmath}
\usepackage{url}
\usepackage{color}
\usepackage{tensor}
\usepackage{mathrsfs}
\usepackage{nicefrac}
\usepackage{tikz-cd}
\usepackage{tikz}
\usetikzlibrary{decorations.markings}
\usetikzlibrary{matrix,arrows}
\usepackage{array}
\usepackage{ulem}
\usepackage{placeins}

\usepackage{pdflscape}

\usepackage[latin1]{inputenc}      
\usepackage[T1]{fontenc}    
\usepackage{ae,aecompl}                          
\usepackage{dsfont}
\usepackage{appendix}
\usepackage{amsfonts,amsmath,latexsym,bbm}
\usepackage{amssymb}
\usepackage{accents}
\usepackage{graphics}
\usepackage{graphicx}
\usepackage{booktabs,longtable}
\usepackage{bm}
\usepackage{dcolumn}
\usepackage{enumitem}
\usepackage{upgreek}
\usepackage{tensor}

\usepackage{tikz}
\usepackage{tikz-cd}
\usepackage{tikz-3dplot}
\usetikzlibrary{decorations.markings}
\usetikzlibrary{matrix,arrows}
\usetikzlibrary{calc}
\usetikzlibrary{shapes}
\usetikzlibrary{positioning}
\usetikzlibrary{fit}

\usepackage{pgfplots}
\pgfplotsset{width=7cm,compat=1.5.1}

\usetikzlibrary{decorations.pathreplacing}
\makeatletter
\def\underbrace#1{%
  \@ifnextchar_{\tikz@@underbrace{#1}}{\tikz@@underbrace{#1}_{}}}
\def\tikz@@underbrace#1_#2{%
  \tikz[baseline=(a.base)] {\node[inner sep=2] (a) {\(#1\)};
  \draw[thick,line cap=round,decorate,decoration={brace,amplitude=4pt}]
    (a.south east) -- node[pos=0.5,below,inner sep=7pt] {\(\scriptstyle #2\)} (a.south west);}}
\makeatother

\usepackage[linktocpage=true]{hyperref}

\makeatletter
\renewcommand*{\eqref}[1]{%
  \hyperref[{#1}]{\textup{\tagform@{\ref*{#1}}}}%
}
\makeatother

\setlist{font=\normalfont\itshape} 

\renewcommand{\vec}[1]{\boldsymbol{#1}}
\newcommand{\tsr}[1]{\overset\leftrightarrow{#1}}
\newcommand{\ext}{_{\rm ext}}
\newcommand{\tot}{_{\rm tot}}
\newcommand{\ind}{_{\rm ind}}

\newcommand{\h}{\hspace{1pt}}
\newcommand{\hh}{\hspace{0.5pt}}
\newcommand{\mh}{\hspace{-1pt}}

\newcommand{\de}{\mathrm d}

\newcommand{\lar}[1]{\textnormal{\mbox{\large $#1$}}}
\renewcommand{\i}{\mathrm i}

\definecolor{oldgray}{gray}{0.4}

\newcommand{\raisemath}[1]{\mathpalette{\raisemith{#1}}}
\newcommand{\raisemith}[3]{\raisebox{#1}{$#2#3$}}

\newcommand{\EE}{_{\raisemath{-2pt}{EE}}}
\newcommand{\EB}{_{\raisemath{-2pt}{EB}}}
\newcommand{\BE}{_{\raisemath{-2pt}{BE}}}
\newcommand{\BB}{_{\raisemath{-2pt}{BB}}}

\newcommand{\T}{_{\mathrm T}}
\renewcommand{\L}{_{\mathrm L}}

\DeclareMathAlphabet{\mathbbmsl}{U}{bbm}{m}{sl}

\numberwithin{equation}{section}

\begin{document}

\begin{frontmatter}



\title{Linear electromagnetic wave equations in materials}


\author[freiberg]{R.~Starke}
\ead{Ronald.Starke@physik.tu-freiberg.de}

\author[heidelberg,aachen]{G.A.H.~Schober\corref{cor1}}
\ead{schober@physik.rwth-aachen.de}

\cortext[cor1]{Corresponding author.}

\address[freiberg]{Institute for Theoretical Physics, TU Bergakademie Freiberg, \\ Leipziger Stra\ss e 23, 09596 Freiberg, Germany \vspace{0.1cm}}
\address[heidelberg]{Institute for Theoretical Physics, Heidelberg University, \\ Philosophenweg 19, 69120 Heidelberg, Germany \vspace{0.1cm}}
\address[aachen]{Institute for Theoretical Solid State Physics, RWTH Aachen University, Otto-Blumenthal-Stra\ss e 26, 52074 Aachen}

\begin{abstract}
After a short review of microscopic electrodynamics in materials, we investigate
the relation of the microscopic dielectric tensor to the current response tensor and to
the full electromagnetic Green function.
Subsequently, we give a systematic overview of microscopic electromagnetic wave equations in materials,
which can be formulated in terms of the microscopic dielectric tensor.
\end{abstract}


\end{frontmatter}



\newpage
\tableofcontents
\newpage

\section{Introduction}\label{sec:intro}

The derivation of linear electromagnetic wave equations {\it in materials} is both conceptually and practically of a fundamental importance.
On the~practical side, it is evident that wave equations in materials form the basis for the description of pulse propagation in media,
a topic which is even part of the standard textbook literature (see e.g.~\cite[Ch.~7]{Jackson} or \cite[Ch.~18 and Ch.~19]{Zangwill}).
On the theoretical side, wave equations in media allow one to relate the {\it speed of light in materials} and hence the {\it refractive index}
to electromagnetic response properties such as the {\it dielectric function} \cite{Strinati,Boeij,Bussi,Romaniello,Berger,Spitaler,Lee,Laszlo,Kawai,Nath,LiLi,Hassanien16diel,Hassanien16opt, Schwalbe,Draxl16}.

Although this is in principle incontrovertible, it is less evident that these problems require a re-investigation from the viewpoint of modern {\it ab initio materials physics}. This concerns, in particular, the standard wave equation in media and the standard relation for the refractive index (see the paradigmatic discussion in Ref.~\cite{Refr}).
In this article, we resume this line of research by a systematic discussion of linear electromagnetic wave equations in materials.
For this purpose, we rely on a microscopic approach to electrodynamics of media, which is well-established in first-principles materials science and in
plasma physics (see \cite{Melrose,Kaxiras,Bruus,Giuliani,Melrose1Book,Altland,Fliessbach,Melrose2Book} for modern textbooks), and which is axiomatized by the Functional Approach to electrodynamics of media \mbox{(see Refs.~\cite{Refr,ED1,ED2,EDOhm,EffWW,EDLor,EDFullGF}).}

Concretely, this article is organized as follows: After assembling in Sct.~\ref{Sec:Review} a number of essential facts and useful formulae from
classical electrodynamics, we proceed in Sct.~\ref{Sec:MicroscopicED} with a review of microscopic electrodynamics of materials. This then allows us to derive electromagnetic wave equations in materials systematically in Sct.~\ref{Sec:WE}.
Finally, Sct.~\ref{sec_iso} investigates these findings in the practically important case of isotropic materials.

\section{Basic notions of classical electrodynamics} \label{Sec:Review}

\subsection{Cartesian projector formalism} \label{sec_proj}

We begin by introducing some technical definitions, which will be frequently used throughout  this article. 
The {\itshape Cartesian longitudinal} and {\itshape transverse projectors} are defined as operators acting on the three-dimensional space \cite[\S\,2.1]{ED1}, i.e.,
\begin{align}
 (P_{\mathrm L} )_{ij}(\vec k) & = \frac{k_i \hh k_j}{|\vec k|^2} \,, \label{eq_defPL} \\[2pt]
 (P_{\mathrm T} )_{ij}(\vec k) & = \delta_{ij} - \frac{k_i \hh k_j}{|\vec k|^2} \,. \label{eq_defPT}
\end{align}
With these, any $(3 \times 3)$ {\itshape Cartesian tensor} $C_{ij}(\vec k)$ can be uni\-{}quely decomposed into four contributions,
\begin{equation}
\tsr C(\vec k) = \tsr C_{\rm LL}(\vec k) + \tsr C_{\rm LT}(\vec k) + \tsr C_{\rm TL}(\vec k) + \tsr C_{\rm TT}(\vec k) \,,
\end{equation}
which are respectively given by
\begin{align}
 \tsr C_{\rm LL}(\vec k) = \tsr P\L(\vec k) \, \tsr C(\vec k) \h \tsr P\L(\vec k) \,, \label{defz_1} \\[3pt]
 \tsr C_{\rm LT}(\vec k) = \tsr P\L(\vec k) \, \tsr C(\vec k) \h \tsr P\T(\vec k) \,, \label{defz_2} \\[3pt]
 \tsr C_{\rm TL}(\vec k) = \tsr P\T(\vec k) \, \tsr C(\vec k) \h \tsr P\L(\vec k) \,, \label{defz_3} \\[3pt]
 \tsr C_{\rm TT}(\vec k) = \tsr P\T(\vec k) \, \tsr C(\vec k) \h \tsr P\T(\vec k) \,. \label{defz_4}
\end{align}
For Cartesian response tensors {\itshape in the isotropic limit,} however, one can prove a simpler decomposition which reads (see \cite[Appendix D.1]{EffWW})
\begin{equation}
 \tsr C(\vec k) = C\L(\vec k) \h \tsr P\L(\vec k) + C\T(\vec k) \h \tsr P\T(\vec k) \,,
\end{equation}
with the {\itshape scalar} longitudinal and transverse coefficient functions $C\L(\vec k)$ and $C\T(\vec k)$. This means, the four contributions defined by Eqs.~\eqref{defz_1}--\eqref{defz_4} are given in terms of these scalar functions by
\begin{align}
 \tsr C_{\rm LL}(\vec k) & = C\L(\vec k) \h \tsr P\L(\vec k) \,, \\[3pt]
 \tsr C_{\rm LT}(\vec k) & = 0 \,, \\[3pt]
 \tsr C_{\rm TL}(\vec k) & = 0 \,, \\[3pt]
 \tsr C_{\rm TT}(\vec k) & = C\T(\vec k) \h \tsr P\T(\vec k) \,.
\end{align}
Conversely, the scalar coefficient functions can be gained back from the original isotropic tensor through
\begin{align} \label{retrieve_long}
 C_{\rm L}(\vec k) & = \frac{\vec k^{\rm T} \, \tsr C(\vec k) \h \vec k}{|\vec k|^2} \,, \\[5pt]
 C_{\rm T}(\vec k) & = \vec e^{\rm T}(\vec k) \, \tsr C(\vec k) \h\hh \vec e(\vec k) \,, \smallskip
\end{align}
where $\vec e(\vec k)$ denotes an arbitrary vector which is transverse and normalized, i.e.~it has the properties that $\vec e(\vec k) \cdot \vec k = 0$ and $\vec e(\vec k) \cdot \vec e(\vec k) = 1$. Finally, one also defines the {\itshape Cartesian transverse rotation operator} as \cite[\S\,2.1]{ED1}
\begin{equation}
 (R_{\rm T})_{ij}(\vec k) = \epsilon_{i\ell j} \h \frac{k_\ell}{|\vec k|} \,.
\end{equation}
The three Cartesian operators $P\L(\vec k)$, $P\T(\vec k)$ and $R\T(\vec k)$ form an algebra, whose multiplication table is given in \cite[Table 1]{ED1}.

\subsection{Free electromagnetic Green function} \label{subsec_ClassED}

\subsubsection{Minkowskian temporal gauge}

As a consequence of the microscopic Maxwell equations, the electromagnetic four-potential, $A^\mu=(\varphi/c,\vec A)^{\rm T}$, 
obeys in terms of the electromagnetic four-current, $j^\mu=(c\rho, \h \vec j)^{\rm T}$, the inhomogeneous {\itshape wave equation} (or {\itshape equation of motion} for the four-potential),
\begin{equation}
(\eta\indices{^\mu_\nu}\Box+\partial^\mu\partial_\nu) \h A^\nu(x)=\mu_0 \h j^\mu(x) \,, \label{fund_wave_eq}
\end{equation}
where $\Box = -\partial^\mu \partial_\mu$ denotes the d'Alembert operator. 
Spelling out spatial and temporal components, we obtain from this the coupled wave equations
\begin{align}
\left(\frac{1}{c^{\,2}}\frac{\partial^2}{\partial t^2}-\Delta\!\right)\!\varphi(\vec x,t)-
\frac{\partial}{\partial t} \mh \left(\frac{1}{c^{\,2}}\frac{\partial \varphi(\vec x, t)}{\partial t} +\nabla \mh \cdot \mh \vec A(\vec x,t)\right)
&=\frac {\rho(\vec x, t)}{\varepsilon_0} \,,\label{eq_EoMPhi}\\[5pt]
\left(\frac{1}{c^{\,2}}\frac{\partial^2}{\partial t^2}-\Delta\!\right)\!\vec{A}(\vec x,t)+
\nabla \mh \left(\frac{1}{c^{\,2}}\frac{\partial \varphi(\vec x, t)}{\partial t}+\nabla \mh \cdot \mh \vec A(\vec x,t)\right)&=\mu_0 \h \vec{j}(\vec x,t) \,.\label{eq_EoMA}
\end{align}
The solutions of these equations are underdetermined in the sense that with any solution $A^\mu$, the ``gauge-transformed'' four-potential $A^\mu+\partial^\mu \mh f$ (where $f$ is an arbitrary scalar function) yields another solution. In particular, the {\it pure gauges}, 
i.e., the four-potentials of the form $A^\mu=\partial^\mu \mh f$ or equivalently,
\begin{align}
\varphi(\vec x,t)&=-\frac{\partial f(\vec x,t)}{\partial t} \,, \label{eq_puregauge1}\\[5pt]
\vec A(\vec x,t)&=\nabla f(\vec x,t)\,, \label{eq_puregauge2}
\end{align}
solve the corresponding {\it homogeneous} equations of motion. This means, for any scalar function $f$, the ansatz \eqref{eq_puregauge1}--\eqref{eq_puregauge2} solves Eqs.~\eqref{eq_EoMPhi}--\eqref{eq_EoMA} with vanishing sources.

To obtain definite results, the equations of motion therefore have to be
complemented by {\it gauge conditions}. In this article, we will work in the {\it temporal gauge}, hence we set the scalar potential to zero:
\begin{equation}
\varphi(\vec x,t)\equiv 0\,.
\end{equation}
It will turn out later that the temporal gauge is particularly useful for condensed matter applications, as it allows for a formulation of response theory
in terms of three-dimensional, {\it Cartesian} vector and tensor quantities (as opposed to the four-dimensional {\it Minkowskian} formulation). In the temporal gauge, Eqs.~\eqref{eq_EoMPhi}--\eqref{eq_EoMA} simplify to
\begin{align}
-\frac{\partial}{\partial t}\h(\nabla\cdot\vec A(\vec x,t))&=\frac 1 {\varepsilon_0} \h \rho(\vec x, t)\,,\label{eq_EoMPhi1}\\ 
\left(\frac{1}{c^{\,2}}\frac{\partial^2}{\partial t^2}-\Delta\right)\!\vec{A}(\vec x,t)+
\nabla\mh\left(\nabla\cdot\vec A(\vec x,t)\right)&=\mu_0 \h \vec j(\vec x,t)\,.\label{eq_EoMA1}
\end{align}
In the Fourier domain, this is equivalent to
\begin{align}
 -\omega \h \vec k\cdot\vec A(\vec k,\omega)&=\frac 1 {\varepsilon_0} \h \rho(\vec k, \omega)\,,\label{eq_EoMPhiFourier}\\
 \left(-\frac{\omega^2}{c^2}+|\vec k|^2\right) \! \vec A(\vec k,\omega)-\vec k \h (\vec k\cdot\vec A)(\vec k,\omega)&=\mu_0 \h \vec j(\vec k,\omega)\,.\label{eq_EoMAFourier}
\end{align}
Eliminating the divergence term in the second equation by means of the first equation, we see that the vector potential obeys the equation of motion
\begin{equation}
\left(-\frac{\omega^2}{c^2}+|\vec k|^2\right) \! \vec A(\vec k,\omega)=\mu_0 \h \bigg(\vec j(\vec k,\omega)-\frac{c\vec k}{\omega} \, c\rho(\vec k,\omega)\bigg)\,. \label{eq_EoMA3}
\end{equation}
This can be solved directly as
\begin{equation}
\vec A(\vec k,\omega)=\mathbbmsl D_0(\vec k,\omega) \h \bigg(\vec j(\vec k,\omega)-\frac{c\vec k}{\omega} \, c\rho(\vec k,\omega)\bigg)\,, \label{eq_Ajrho}
\end{equation}
where \smallskip
\begin{equation}
\mathbbmsl D_0(\vec k, \omega) = \frac{\mu_0}{-\omega^2/c^2+ |\vec k|^2} = -\frac{1}{\varepsilon_0 \h \omega^2} \h \frac{1}{1 - c^2|\vec k|^2/\omega^2} \smallskip \vspace{2pt}
\end{equation}
is the scalar Green function of the d'Alembert operator in Fourier space. \linebreak

\pagebreak \noindent
Now, writing Eq.~\eqref{eq_Ajrho} in the Minkowskian form as
\begin{equation}
A^\mu(k)=(D_0)\indices{^\mu_\nu}(k) \h j^\nu(k)\,, \label{eq_introD_0}
\end{equation}
we read off that the free electromagnetic Green function in the temporal gauge is given by 
\begin{equation} \label{eq_weylgreen}
 (D_0)\indices{^\mu_\nu}(\vec k, \omega) = \mathbbmsl D_0(\vec k, \omega) \left( \begin{array}{cc} 0 & 0 \\[5pt]
 -c\vec k / \omega & \tsr 1
 \end{array} \right).
\end{equation}
This version of the electromagnetic Green function in the temporal gauge therefore still gives a Minkowskian, i.e., four-dimensional formulation of the equation of motion (involving both charge and current densities). Consequently, we will refer to Eq.~\eqref{eq_weylgreen} as the free electromagnetic Green function in the {\itshape Minkowskian temporal gauge.} Next, we will perform the transition to a purely Cartesian formulation.

\subsubsection{Cartesian temporal gauge} \label{sec_cart_form}

Within the temporal gauge, the complete information about the
four-potential is encapsulated in its spatial part, the vector potential. Moreover, 
the components of the four-current are also not independent of each other. Instead, they fulfill the continuity equation which reads in the Fourier domain
\begin{equation} \label{cont_eq_covar}
 k_\mu \h\hh j^\mu(k) = 0 \,,
\end{equation}
or equivalently, \smallskip
\begin{equation}
\omega \h\hh \rho(\vec k,\omega)=\vec k\cdot\vec j(\vec k,\omega) \equiv \vec k^{\rm T} \vec j(\vec k, \omega) \,. \label{eq_continuity} \smallskip
\end{equation}
Using this to eliminate the charge density from the equation of motion of the vector potential, Eq.~ \eqref{eq_EoMA3}, we obtain
\begin{equation} \label{wave_A_equiv}
\bigg({-\frac{\omega^2}{c^2}+|\vec k|^2}\bigg)\vec A(\vec k,\omega)=\mu_0\left(\tsr 1-\frac{c^2 \hh \vec k \hh\vec k^{\rm T}}{\omega^2} \h \right)\vec j(\vec k,\omega) \,.
\end{equation}
Consequently, the vector potential is given in terms of the current density as
\begin{equation}  \label{eq_EoMASol_prev}
\vec A(\vec k,\omega)=\mathbbmsl D_0(\vec k,\omega)\left(\tsr 1-\frac{c^2|\vec k|^2}{\omega^2} \h \tsr P\L(\vec k)\right)\vec j(\vec k,\omega)\,.
\end{equation}
This equation can be derived even more directly by using that the equations of motion \eqref{eq_EoMPhiFourier}--\eqref{eq_EoMAFourier} are not independent
of each other. Instead, the first equation follows from the second one by multiplying through with $\vec k^{\rm T}$ and using the continuity equation.
Consequently, Eq.~\eqref{eq_EoMPhiFourier} can be discarded, and hence the complete dynamics is already contained in Eq.~\eqref{eq_EoMAFourier} which can be written equivalently as
\begin{equation}
\bigg({-\frac{\omega^2}{c^2}+|\vec k|^2 \h \tsr P\T(\vec k)}\bigg) \vec A(\vec k,\omega)=\mu_0 \h \vec j(\vec k,\omega)\,. \label{eq_EoMA2}
\end{equation}
Inverting this equation and using the identity
\begin{equation} \label{iden}
\bigg({-\frac{\omega^2}{c^2}+|\vec k|^2 \h \tsr P\T(\vec k)}\bigg)^{\!\!-1} = \bigg({-\frac{\omega^2}{c^2} + |\vec k|^2} \bigg)^{\!\!-1} \, \bigg( \tsr 1 - \frac{c^2 |\vec k|^2}{\omega^2} \h \tsr P_{\rm L}(\vec k) \bigg)
\end{equation}
yields again the desired Eq.~\eqref{eq_EoMASol_prev}.

Thus, we have shown that the vector potential in the temporal gauge can be expressed entirely in terms of the spatial current density by means of the Cartesian tensor equation 
\begin{equation}\label{eq_EoMASol}
\vec A(\vec k,\omega)=\tsr D_0(\vec k,\omega) \, \vec j(\vec k,\omega)\,,
\end{equation}
whereby we have introduced the {\itshape free Cartesian Green function}
\begin{equation}\label{eq_GFtempGauge}
\tsr D_0(\vec k,\omega)=\mathbbmsl D_0(\vec k,\omega)\left(\tsr 1-\frac{c^2|\vec k|^2}{\omega^2} \h \tsr P\L(\vec k)\right).
\end{equation}
By Eq.~\eqref{iden}, its inverse is given by
\begin{equation}
(\tsr D_0 )^{-1}(\vec k,\omega) = \frac{1}{\mu_0} \left( -\frac{\omega^2}{c^2} + |\vec k|^2 \tsr P\T(\vec k) \right). \label{eq_InvCartGF}
\end{equation}
For later purposes, we further rewrite Eq.~\eqref{eq_GFtempGauge} as
\begin{equation} \label{freeCart_decom}
 \tsr D_0(\vec k,\omega) = D_{0, \hh \rm L}(\vec k, \omega) \h  \tsr P\L(\vec k) + D_{0, \hh \rm T}(\vec k, \omega) \h \tsr P\T(\vec k) \,,
\end{equation}
where the longitudinal and transverse components are given by
\begin{align}
 D_{0, \hh \rm L}(\vec k, \omega) & = \mathbbmsl D_0(\vec k, \omega) \h \bigg( 1 - \frac{c^2 |\vec k|^2}{\omega^2} \bigg) = -\frac1 { \varepsilon_0 \h \omega^2} \,, \label{freeCart_long} \\[5pt]
 D_{0, \hh \rm T}(\vec k, \omega) & = \mathbbmsl D_0(\vec k, \omega) \,. \label{freeCart_trans}
\end{align}
These formulae for the free Green function are generally valid, which stands in contrast to the full Green function having an analogous representation only in the isotropic limit (see \S\,\ref{longtranseps}).

We conclude that in the temporal gauge, the four-dimensional electromagnetic Green function can also be chosen as
\begin{equation} \label{eq_weylgreen1}
 (D_0)\indices{^\mu_\nu}(\vec k, \omega) = \left( \! 
 \begin{array}{cc} 0 & 0 \\[3pt]
 0 & \tsr D_0(\vec k, \omega)
 \end{array} \! \right),
\end{equation}
with the Cartesian Green function given by Eq.~\eqref{eq_GFtempGauge}. As this expression leads to a Cartesian, i.e., three-dimensional formulation of the equation of motion, we refer to it as the free electromagnetic Green function in the {\itshape Cartesian temporal gauge.} Finally, we remark that both the Minkowskian and the Cartesian forms, Eqs.~\eqref{eq_weylgreen} and \eqref{eq_weylgreen1}, are consistent with the most general expression of the free electromagnetic Green function derived in Ref.~\cite[\S\,3.3]{ED1}. This is shown explicitly in Ref.~\cite[\S\,3.3]{EDFullGF}.

\subsection{Electric and magnetic solution generators}

The equations of motion for the electric and magnetic fields can be deduced directly from the three-dimensional equation of motion for the vector-potential, Eq.~\eqref{eq_EoMA2}. Using the representation of the fields in terms of the potentials in the temporal gauge,
\begin{align}
\vec E(\vec k,\omega)&=\i\omega\vec A(\vec k,\omega)\,,\label{eq_EA}\\[5pt]
\vec B(\vec k,\omega)&=\i\vec k\times\vec A(\vec k,\omega)\,,
\end{align}
it follows that the wave equations for the electric and magnetic fields read
\begin{align}
\left(-\frac{\omega^2}{c^2}+|\vec k|^2-\vec k\vec k^{\rm T}\right) \! \vec E(\vec k,\omega)&=\mu_0 \h \i\omega \h \vec j(\vec k,\omega)\,, \label{eq_EoME} \\[3pt]
\left(-\frac{\omega^2}{c^2}+|\vec k|^2\right) \! \vec B(\vec k,\omega)&=\mu_0 \h \i\vec k\times\vec j(\vec k,\omega)\,. \label{eq_EoMB}
\end{align}
For the second equation, we have used that by Ref.~\cite[Table 1]{ED1},
\begin{equation}
 \vec k \times \tsr P\T(\vec k) \vec A \h = \h  |\vec k| \h \tsr R\T(\vec k) \hh \tsr P\T(\vec k) \vec A \h = \h  |\vec k| \h \tsr R\T(\vec k) \vec A \h = \h  \vec k \times \vec A \,.
\end{equation}
Finally, using Eqs.~\eqref{eq_EoMPhiFourier} and \eqref{eq_EA} we obtain the constraint
\begin{equation}
\i\vec k\cdot\vec E(\vec k,\omega)=\frac 1 {\varepsilon_0} \h \rho(\vec k, \omega) \,, \label{eq_Gauss}
\end{equation}
which coincides with Gauss' law, and thus the equation of motion for the electric field reverts to
\begin{equation}\label{eq_FundWaveSourFourier}
\left(-\frac{\omega^2}{c^2}+|\vec k|^2\right) \!\vec E(\vec k,\omega)=-\frac{1}{\varepsilon_0} \h \i\vec k\hh \rho(\vec k,\omega) + \mu_0 \h \i\omega \hh \vec j(\vec k,\omega)\,,
\end{equation}
which is the standard wave equation for the electric field. 

The equations  \eqref{eq_EoMA2}, \eqref{eq_EoME} and \eqref{eq_EoMB} can be solved explicitly
for $\vec A$, $\vec E$ and $\vec B$ by means of elementary algebraic operations. In the case of the vector potential, the solution has already been given in Eqs.~\eqref{eq_EoMASol}--\eqref{eq_GFtempGauge}.
Similarly, we introduce the {\it solution generators} for the electric and magnetic fields by means of the defining equations
\begin{align}
\vec E(\vec k,\omega)&=\frac{1}{\i\omega\varepsilon_0} \h \tsr{\mathbbmsl E}(\vec k,\omega)\,\vec j(\vec k,\omega)\,, \label{el_sol} \\[5pt]
c \h \vec B(\vec k,\omega)&=\frac{1}{\i\omega\varepsilon_0} \h \tsr{\mathbbmsl B}(\vec k,\omega)\,\vec j(\vec k,\omega)\,. \label{mag_sol}
\end{align}
These solution generators are given explicitly by the dimensionless expressions (compare also \cite[Eqs.~(4.9)--(4.12)]{ED1})
\begin{align}
\tsr{\mathbbmsl E}(\vec k,\omega)&=-\varepsilon_0 \h\hh \omega^2\tsr{D}_0(\vec k,\omega)\,, \label{el_sol_exp} \\[5pt]
\tsr{\mathbbmsl B}(\vec k,\omega)&=- \varepsilon_0 \h\hh \omega \, c|\vec k| \h \tsr R\T(\vec k)\tsr{D}_0(\vec k,\omega)\,,
\end{align}
which will be used for deriving the Universal Response Relations in~\S\,\ref{subsec_URR}.

\subsection{Total functional derivatives} \label{subsec_total_der}

The free electromagnetic Green function introduced in Eq.~\eqref{eq_introD_0} can be characterized as the functional derivative \cite[\S\,5.2]{ED1}
\begin{equation}
 (D_0)\indices{^\mu_\nu}(x - x') = \frac{\delta A^\mu(x)}{\delta j^\nu(x')} \,. \label{eq_DefFreeGF}
\end{equation}
In Fourier space, this is equivalent to
\begin{equation}
 \frac{\delta A^\mu(k)}{\delta j^\nu(k')} = (D_0)\indices{^\mu_\nu}(k) \h \delta^4(k - k') \,,
\end{equation}
which we will often abbreviate as
\begin{equation}
 \frac{\delta A^\mu(k)}{\delta j^\nu(k)} = (D_0)\indices{^\mu_\nu}(k) \,.
\end{equation}
From Eq.~\eqref{eq_Ajrho} we obtain the equalities
\begin{align}
\frac{\delta\vec A(\vec k,\omega)}{\delta\vec j(\vec k,\omega)}&=\mathbbmsl D_0(\vec k,\omega) \h \tsr 1 \,, \label{zwischen_5} \\[5pt]
\frac{\delta\vec A(\vec k,\omega)}{\delta\rho(\vec k,\omega)}&=-\frac{c^2\vec k}{\omega} \h \mathbbmsl D_0(\vec k,\omega)\,. \label{zwischen_6}
\end{align}
With these, we now show that the Cartesian Green function can be characterized as the {\it total functional derivative} of the vector potential with respect to the spatial current, i.e.,
\begin{equation}\label{eq_totDerCartGF}
\tsr D_0(\vec k,\omega)=\frac{\de\vec A(\vec k,\omega)}{\de\vec j(\vec k,\omega)} \,,
\end{equation}
which means in components,
\begin{equation}
 (D_0)_{i\ell}(\vec k, \omega) = \frac{\de A_i(\vec k, \omega)}{\de j_\ell(\vec k, \omega)} \,.
\end{equation}
Here, the total derivative is defined as
\begin{equation} \label{def_total_der_curr}
 \frac{\de}{\de j_\ell(\vec k,\omega)}= \frac{\delta}{\delta j_\ell(\vec k,\omega)}+\frac{\delta\rho(\vec k,\omega)}{\delta j_\ell(\vec k,\omega)} \h \frac{\delta}{\delta\rho(\vec k,\omega)}\,,
\end{equation}
with the derivative of the charge density with respect to the current density being defined through the continuity equation as
\begin{equation}\label{eq_delrho_delj}
\frac{\delta\rho(\vec k,\omega)}{\delta j_\ell(\vec k,\omega)}=\frac{k_\ell}{\omega}\,. 
\end{equation}
In fact, by substituting this last formula into Eq.~\eqref{def_total_der_curr} and using the expressions \eqref{zwischen_5}--\eqref{zwischen_6} for the partial derivatives, one shows directly that Eq.~\eqref{eq_totDerCartGF} is equivalent to the defining Eq.~\eqref{eq_GFtempGauge}. Similarly, the solution generators 
introduced in Eqs.~\eqref{el_sol}--\eqref{mag_sol} can be characterized as
\begin{align}
\frac{\de\vec E(\vec k,\omega)}{\de\vec j(\vec k,\omega)}&=\frac{1}{\i\omega\varepsilon_0} \h \tsr{\mathbbmsl E}(\vec k,\omega)\,,\\[5pt]
c\,\frac{\de\vec B(\vec k,\omega)}{\de\vec j(\vec k,\omega)}&=\frac{1}{\i\omega\varepsilon_0} \h \tsr{\mathbbmsl B}(\vec k,\omega)\,,
\end{align}
i.e., as total functional derivatives with respect to the current density.

Next, a total functional derivative with respect to the vector potential can be defined analogously to Eq.~\eqref{def_total_der_curr} as
\begin{equation} \label{zwischen_7}
\frac{\de}{\de A_\ell(\vec k, \omega)}=\frac{\delta}{\delta A_\ell(\vec k, \omega)} + \frac{\delta\varphi(\vec k, \omega)}{\delta A_\ell(\vec k, \omega)} \h \frac{\delta}{\delta\varphi(\vec k, \omega)}\,,
\end{equation}
provided that we are given a gauge condition in the form $\varphi = \varphi[\vec A]$.
Such a functional dependence of the scalar potential on the vector potential is indeed well-defined for the Lorenz gauge, where $\varphi[\vec A] = (c^2 / \omega) \h \vec k\cdot\vec A$, 
and for the temporal gauge, where $\varphi[\vec A] = 0$ trivially. In particular, in the temporal gauge used in this article, the second term in Eq.~\eqref{zwischen_7} vanishes, and hence in 
this case total and partial functional derivatives coincide.

Finally, total functional derivatives with respect to electric and magnetic fields have already been defined in Ref.~\cite{ED1} as
\begin{align}
 \frac{\de \vec A(\vec k, \omega)}{\de \vec E(\vec k, \omega)} & = \frac{\delta \vec A(\vec k, \omega)}{\delta \vec E(\vec k, \omega)} + \frac{\delta \vec A(\vec k, \omega)}{\delta \vec B(\vec k, \omega)} \h \frac{\delta \vec B(\vec k, \omega)}{\delta \vec E(\vec k, \omega)} \,, \\[5pt]
 \frac{\de \vec A(\vec k, \omega)}{\de \vec B(\vec k, \omega)} & = \frac{\delta \vec A(\vec k, \omega)}{\delta \vec B(\vec k, \omega)} + \frac{\delta \vec A(\vec k, \omega)}{\delta \vec E(\vec k, \omega)} \h \frac{\delta \vec E(\vec k, \omega)}{\delta \vec B(\vec k, \omega)} \,,
\end{align}
where by Faraday's law, the mutual partial derivatives are given by
\begin{align}
 \frac{\delta \vec B(\vec k, \omega)}{\delta \vec E(\vec k, \omega)} & = \frac{|\vec k|}{\omega} \h \tsr R(\vec k) \,, \\[5pt]
 \frac{\delta \vec E(\vec k, \omega)}{\delta \vec B(\vec k, \omega)} & =  - \frac{\omega}{|\vec k|} \h \tsr R(\vec k) \,.
\end{align}
Concretely, the total derivatives of the vector potential in the temporal gauge with respect to the electric and magnetic fields are given by 
\begin{align}
\frac{\de\vec A(\vec k,\omega)}{\de\vec E(\vec k,\omega)}&=\frac{1}{\i \hh \omega} \h \tsr 1\,, \label{eq_totDerAE}\\[5pt]
\frac{\de\vec A(\vec k,\omega)}{\de\vec B(\vec k,\omega)}&=-\frac{1}{\i \hh |\vec k|} \h \tsr R\T(\vec k)\,,\label{eq_totDerAB}
\end{align}

\vspace{2pt} \noindent
as we have shown in Ref.~\cite[\S\,4.2]{ED1}.

\section{Microscopic electrodynamics of materials}\label{Sec:MicroscopicED}

\subsection{Fundamental response tensor}

Microscopic electrodynamics of media is based on the postulated functional dependence of {\itshape induced} field quantities 
on {\itshape external} perturbations (see Ref.~\cite{ED2} for a paradigmatic discussion). In particular, the functional derivative of
an induced quantity with respect to an external perturbation is called a {\it response function}. In the electrodynamic context, it is particularly convenient to consider the induced current density $j^\mu\ind(x)$ as a functional of the external four-potential $A^\nu\ext(x')$, i.e.,
\begin{equation}
j^\mu\ind=j^\mu\ind[A^\nu\ext] \,.
\end{equation}
To linear order in the external field, this implies the Lorentz-covariant expansion \smallskip
\begin{equation}
 j^\mu\ind(x)=\int \! \de^4 x'\,\chi\indices{^\mu_\nu}(x,x') \h A^\nu\ext(x')\,, \smallskip
\end{equation}
where the integral kernel
\begin{equation}
 \chi\indices{^\mu_\nu}(x,x')=\frac{\delta j^\mu\ind(x)}{\delta A^\nu\ext(x')} \label{eq_FundRespTens}
\end{equation}
is called {\it fundamental response tensor} (see \cite[Sct.~5.1]{ED1}). On account of the continuity equation and the gauge-independence of the induced four-current,
it has to obey the constraints
\begin{equation}
\partial_\mu \h \chi\indices{^\mu_\nu}(x,x')= \partial'^{\nu}\chi\indices{^\mu_\nu}(x,x')=0 \,. \label{eq_constraints}
\end{equation}
From this, it follows that the four-dimensional fundamental response tensor can be brought into the following form in Fourier space (see \cite[\S\,4.2]{EDFullGF}):
\begin{equation}\label{generalform1}
\chi^\mu_{~\nu}(\vec k,\omega)=
\left( \!
\begin{array}{rr} -\lar{\frac{c^2}{\omega^2}} \, \vec k^{\rm T} \, \tsr{\chi}(\vec k,\omega)\, \vec k & \lar{\frac{c}{\omega}} \, \vec k^{\rm T} \, \tsr{\chi}(\vec k,\omega)\, \\
[10pt] -\lar{{\frac{c}{\omega}}} \,\h \tsr{\chi}(\vec k,\omega)\, \vec k & \, \tsr{\chi}(\vec k,\omega)\,
\end{array} \right).
\end{equation}
In this formula, we have implicitly assumed the {\itshape homogeneous limit,} by which
any response function in real space depends only on the difference of its two arguments, $\chi(x,x')=\chi(x-x')$,
such that in Fourier space it is given by a function of one argument only, $\chi(k, k')= \delta^4(k - k') \h \chi(k)$. In the sequel, we will always keep to
this assumption.

\subsection{Universal Response Relations}\label{subsec_URR}

The general form of the fundamental response tensor, Eq.~\eqref{generalform1}, implies that  the Cartesian (i.e., $3\times 3$) {\itshape current response tensor,}
\begin{equation}
 \tsr{\chi}(\vec k,\omega)=\frac{\delta\vec j\ind(\vec k,\omega)}{\delta\vec A\ext(\vec k,\omega)}\,,
\end{equation}
already determines uniquely the whole Minkowskian (i.e., $4 \times 4$) fundamental response tensor. Since the latter
determines all linear electromagnetic response properties, we conclude that any linear electromagnetic response function can be expressed analytically
in terms of the Cartesian current response tensor. In particular, this applies to the response
of induced electric or magnetic fields with respect to external electric or magnetic fields. Concretely, by applying the functional chain rule we obtain
\begin{align}
\frac{\de\vec E\ind}{\de\vec E\ext}&=\frac{\de\vec E\ind}{\de\vec j\ind}\h \bigg(\frac{\de\vec j\ind}{\de\vec A\ext}\bigg) \h \frac{\de\vec A\ext}{\de\vec E\ext}\,, \label{URR1.1}\\[5pt]
\frac 1 c \h \frac{\de\vec E\ind}{\de\vec B\ext}&=\frac 1 c \h \frac{\de\vec E\ind}{\de\vec j\ind}\h \bigg(\frac{\de\vec j\ind}{\de\vec A\ext}\bigg) \h \frac{\de\vec A\ext}{\de\vec B\ext}\,,\label{URR1.2}\\[5pt]
c\,\frac{\de\vec B\ind}{\de\vec E\ext}&=c\,\frac{\de\vec B\ind}{\de\vec j\ind}\h \bigg(\frac{\de\vec j\ind}{\de\vec A\ext}\bigg) \h \frac{\de\vec A\ext}{\de\vec E\ext}\,,\label{URR1.3}\\[5pt]
\frac{\de\vec B\ind}{\de\vec B\ext}&=\frac{\de\vec B\ind}{\de\vec j\ind}\h \bigg(\frac{\de\vec j\ind}{\de\vec A\ext}\bigg) \h \frac{\de\vec A\ext}{\de\vec B\ext}\,. \label{URR1.4}
\end{align}
Using results from $\S\,\ref{subsec_total_der}$, this yields the {\it Universal Response Relations} which in the homogeneous limit read
\begin{align}
\tsr\chi\EE(\vec k,\omega)&=-\frac{1}{\varepsilon_0 \h\hh \omega^2} \, \tsr{\mathbbmsl E}(\vec k,\omega) \h \tsr\chi(\vec k,\omega)\,,\\[5pt]
\tsr\chi\EB(\vec k,\omega)&=\frac{1}{\varepsilon_0 \h\hh  \omega \h\hh c|\vec k|} \, \tsr{\mathbbmsl E}(\vec k,\omega) \h \tsr\chi(\vec k,\omega) \h \tsr R\T(\vec k)\,,\\[5pt]
\tsr\chi\BE(\vec k,\omega)&=-\frac{1}{\varepsilon_0 \h\hh \omega^2} \, \tsr{\mathbbmsl B}(\vec k,\omega) \h \tsr\chi(\vec k,\omega)\,,\\[5pt]
\tsr\chi\BB(\vec k,\omega)&=\frac{1}{\varepsilon_0 \h\hh \omega \h\hh c|\vec k|} \, \tsr{\mathbbmsl B}(\vec k,\omega) \h \tsr\chi(\vec k,\omega) \h \tsr R\T(\vec k)\,,
\end{align}
where $\chi\EE=\de\vec E\ind/\de\vec E\ext$\h, etc. Furthermore, with the standard relation between the current response tensor and the conductivity tensor,
\begin{equation} \label{chisigma}
\tsr\chi(\vec k,\omega)=\i\omega\tsr\sigma(\vec k,\omega)\,, 
\end{equation}
and with the Cartesian Green function defined in Eq.~\eqref{eq_GFtempGauge}, the Universal Response Relations can also be written as
\begin{align}
\tsr\chi\EE(\vec k,\omega)&=\i\omega \h \tsr{D}_0(\vec k,\omega) \h \tsr\sigma(\vec k,\omega)\,, \label{reURR1}\\[5pt]
\tsr\chi\EB(\vec k,\omega)&=\left( \i\omega\h\tsr{D}_0(\vec k,\omega) \h\tsr\sigma(\vec k,\omega)\right) \mh \frac{(-\omega)}{c|\vec k|}\h\tsr R\T(\vec k)\,,\\[4pt]
\tsr\chi\BE(\vec k,\omega)&= \frac{c|\vec k|}{\omega} \h \tsr R\T(\vec k) \mh \left(\i\omega\h \tsr D_0(\vec k,\omega) \h \tsr\sigma(\vec k,\omega)\right),\\[8pt]
\tsr\chi\BB(\vec k,\omega)&=-\tsr R\T(\vec k) \mh \left(\i\omega\h \tsr D_0(\vec k,\omega)\h \tsr\sigma(\vec k,\omega)\right)\mh \tsr R\T(\vec k)\,.
\end{align}
These equations can be derived even more directly from the chain rules
\begin{align}
\tsr\chi\EE&=\frac{\de \vec E\ind}{\de \vec E\ext} \,, \\[5pt]
\tsr\chi\EB&= \frac 1 c \h \bigg( \frac{\de \vec E\ind}{\de \vec E\ext} \bigg) \h \frac{\de \vec E\ext}{\de \vec B\ext} \,, \\[5pt]
\tsr\chi\BE&= c \, \frac{\de \vec B\ind}{\de \vec E\ind} \h \bigg( \frac{\de \vec E\ind}{\de \vec E\ext} \bigg) \,,\\[5pt]
\tsr\chi\BB&= \frac{\de \vec B\ind}{\de \vec E\ind} \h \bigg( \frac{\de \vec E\ind}{\de \vec E\ext} \bigg) \h \frac{\de \vec E\ext}{\de \vec B\ext} \,.
\end{align}
In Ref.~\cite{ED1}, it is shown systematically that all standard relations between electromagnetic response functions can be recovered from the Universal Response Relations in suitable limiting cases.
We conclude, in particular, that all physical response functions can be calculated explicitly in terms of the conductivity tensor.

\subsection{Full electromagnetic Green function}\label{subsec_FGF}

Yet another response tensor of fundamental importance is the {\it full} electromagnetic Green function.
This is the direct counterpart of the {\it free} electromagnetic Green function, which by Eq.~\eqref{eq_DefFreeGF} can be characterized as the \linebreak functional derivative of the (external) four-potential with respect to the \linebreak (external) four-current.
Correspondingly, the {\it full} Green function is defined as the functional derivative of the {\itshape total} four-potential with respect to the {\itshape external} four-current, i.e.,
\begin{equation}
 D\indices{^\mu_\nu}(x,x')=\frac{\delta A^\mu\tot(x)}{\delta j^\nu\ext(x')}\,.
\end{equation}
The full and the free electromagnetic Green function are related through the Dyson equation, \smallskip
\begin{equation} \label{dyson_eq}
D=D_0+D_0 \h \widetilde\chi \h D\,, \smallskip
\end{equation}
where $\widetilde\chi$ denotes the {\it proper} fundamental response tensor, 
\begin{equation}
 \widetilde\chi\indices{^\mu_\nu}(x,x')=\frac{\delta j^\mu\ind(x)}{\delta A^\nu\tot(x')} \,.
\end{equation}
The latter is in turn connected to the ``direct'' fundamental response tensor defined in Eq.~\eqref{eq_FundRespTens} by a Dyson-type equation:
\begin{equation} \label{rel_dir_prop}
\chi=\widetilde\chi+\widetilde\chi \h D_0 \h \chi\,.
\end{equation}
Now, if we choose the free electromagnetic Green function in the Cartesian temporal gauge (see  Eq.~\eqref{eq_weylgreen1}), then one can show (see Ref.~\cite[\S\,5.3]{EDFullGF}) that the full Green function has the same form as the free Green function, i.e., 
\begin{equation} \label{eq_weylgreen1full}
 D\indices{^\mu_\nu}(\vec k, \omega) = \left( \!
 \begin{array}{cc} 0 & 0 \\[3pt]
 0 & \tsr D(\vec k, \omega)
 \end{array} \! \right).
\end{equation}
Furthermore, the {\itshape full Cartesian Green function}, which is defined as the spatial part of the above tensor, obeys a Dyson equation of the Cartesian form,
\begin{equation}
\tsr D=\tsr D_0 +\tsr D_0 \h \tsr{\widetilde\chi} \h \tsr D\,, \label{eq_CartDyson}
\end{equation}
which is analogous to the fundamental Dyson equation \eqref{dyson_eq}. Similarly, as shown in Ref.~\cite[\S\,4.3]{EDFullGF}, the Dyson-type equation \eqref{rel_dir_prop} implies the corresponding Cartesian equation,
\begin{equation} \label{Cart_dyson_1}
 \tsr \chi = \tsr{\widetilde \chi} + \tsr{\widetilde \chi} \h \tsr D_0 \h \tsr \chi \,.
\end{equation}
For later purposes, we provide yet two equivalent formulations of Eq.~\eqref{eq_CartDyson}: First, the inverse of the full Cartesian Green function is given by 
\begin{equation}
 \tsr D{}^{-1} = (\tsr D_0)^{-1} - \tsr{\widetilde \chi} \,. \label{eq_CartDyson_2}
\end{equation}
Secondly, by multiplying this equation through with the free Green function, 
we obtain the identity \smallskip
\begin{equation}
 \tsr D_0 \h \tsr D{}^{-1} = \tsr 1  - \tsr D_0 \h \tsr{\widetilde \chi} \,,\smallskip \label{eq_CartDyson_1}
\end{equation}
which will be used in the next subsection.

Finally, with the above definitions, the total four-potential in the tempo\-{}ral gauge can be expanded to linear order in the external current as
\begin{equation}
\vec A\tot(\vec k,\omega)=\tsr D(\vec k,\omega) \, \vec j\ext(\vec k,\omega)\,, \label{eq_ADjext}
\end{equation}
which is the direct material analogon of Eq.~\eqref{eq_EoMASol} valid ``in vacuo''. In other words, the {\itshape full} Green function relates the total vector potential (which is the sum of the external vector potential and the vector potential induced by the material) in terms of the external current, whereas the {\itshape free} Green function relates each vector potential to its own sources. Correspondingly, exactly as the free Cartesian Green function (see Eq.~\eqref{eq_totDerCartGF}), 
the full Cartesian Green function can be characterized as the total functional derivative with respect to spatial current density (see Eq.~\eqref{def_total_der_curr}),
\begin{equation}\label{eq_totDerFullCartGF}
\tsr D(\vec k,\omega)=\frac{\de\vec A\tot(\vec k,\omega)}{\de\vec j\ext(\vec k,\omega)} \,.
\end{equation}
As we will show next, the full Cartesian Green function is closely related to the dielectric tensor.

\subsection{Dielectric tensor}\label{subsec_DiTens}

Within {\itshape ab initio} materials physics, the dielectric tensor is defined as the functional derivative of the external with respect to 
the total electric field (see e.g.~\cite{Kaxiras,Martin,SchafWegener}), i.e.,
\begin{equation}
\tsr\varepsilon_{\rm r}(x, x') = \frac{\de\vec E_{\rm ext}(x)}{\de\vec E_{\rm tot}(x')} \,. \label{eq_FundDef}
\end{equation}
In the homogeneous limit, this is equivalent to
\begin{equation}
\tsr\varepsilon_{\rm r}(\vec k,\omega)=\frac{\de\vec E\ext(\vec k,\omega)}{\de\vec E\tot(\vec k,\omega)}\,. \label{eq_DefDiTens}
\end{equation}
Within the temporal gauge, we can use the identity $\vec E_{\rm ext} = \i\omega \vec A_{\rm ext}$ and the total derivative of the vector potential with respect to the electric field given by Eq.~\eqref{eq_totDerAE} to obtain the equivalent characterization
\begin{equation}
\tsr\varepsilon_{\rm r}(\vec k,\omega)=\frac{\de\vec A\ext(\vec k,\omega)}{\de\vec A\tot(\vec k,\omega)}\,. \label{eq_Altnerative}
\end{equation}
This can further be rewritten as 
\begin{equation} \label{zwischen_3}
\tsr\varepsilon_{\rm r}\h = \h \frac{\de(\vec A\tot-\vec A\ind)}{\de\vec A\tot} \h = \h  \tsr 1-\frac{\de\vec A\ind}{\de\vec A\tot} \h = \h \tsr 1-\frac{\de\vec A\ind}{\de\vec j\ind}\frac{\de\vec j\ind}{\de\vec A\tot}\,,
\end{equation}
from which we obtain
\begin{equation}
\tsr\varepsilon_{\rm r}(\vec k,\omega)=\tsr 1 - \tsr D_0(\vec k,\omega) \h \tsr{\widetilde\chi}(\vec k,\omega)\,,\label{eq_eps}
\end{equation}
where we have used the identity of total and partial derivatives in the temporal gauge (see \S\,\ref{subsec_total_der}). An alternative derivation of this relation can be given as follows: We have
\begin{equation}
\tsr \varepsilon_{\rm r} \h = \h \tsr 1 - \frac{\de \vec E_{\rm ind}}{\de \vec E_{\rm tot}} \h \equiv \h \tsr 1 - \tsr{\widetilde \chi}{}_{EE} \,,
\end{equation}
which follows from a similar logic as Eq.~\eqref{zwischen_3}. Using the Universal Response Relation \eqref{reURR1} and the equality \eqref{chisigma}---which both hold also for the respective proper response functions---we arrive again at
\begin{equation}
 \tsr\varepsilon_{\rm r} \h = \h  \tsr 1 - \i\omega \h \tsr D_0 \h \tsr{\widetilde \sigma} \h = \h  \tsr 1 - \tsr D_0 \h \tsr{\widetilde \chi} \,.
\end{equation}
Similarly, one shows the relation for the inverse dielectric tensor,
\begin{equation}
(\tsr\varepsilon_{\rm r})^{-1}(\vec k,\omega)=\tsr 1 + \tsr D_0(\vec k,\omega) \h \tsr{\chi}(\vec k,\omega)\,. \label{eq_epsinv}
\end{equation}
Finally, using once more the formalism of total functional derivatives, we now derive the connection between the dielectric tensor and the full Cartesian Green function.
We start from
\begin{equation}
\frac{\de\vec A\tot}{\de\vec j\ext}=\frac{\de\vec A\tot}{\de\vec A\ext} \h \frac{\de\vec A\ext}{\de\vec j\ext}\,,
\end{equation}
which directly translates into the result
\begin{equation}\label{eq_DvsEps}
\tsr D(\vec k,\omega) = (\tsr\varepsilon_{\rm r})^{-1}(\vec k,\omega) \h \tsr D_0(\vec k,\omega)\,.
\end{equation}
This fundamental relation also follows from Eq.~\eqref{eq_eps} by using the Cartesian Dyson equation in the form \eqref{eq_CartDyson_1}. 
It shows that the full Cartesian Green function essentially coincides with the inverse dielectric tensor. Similarly, the 
functional chain rule \smallskip
\begin{equation}
 \frac{\de \vec j_{\rm ind}}{\de \vec A\tot} = \frac{\de \vec j_{\rm ind}}{\de \vec A\ext}\h\frac{\de \vec A\ext}{\de \vec A\tot} \smallskip
\end{equation}
translates into the relation
\begin{equation} \label{tildechichieps}
 \tsr{\widetilde \chi}(\vec k, \omega) = \tsr \chi(\vec k, \omega) \h \tsr\varepsilon_{\rm r}(\vec k, \omega) \,,
\end{equation}
which also follows from Eq.~\eqref{eq_epsinv} and the Cartesian Dyson equation \eqref{Cart_dyson_1}.

\section{Linear wave equations in materials}\label{Sec:WE}

\subsection{Homogeneous wave equations}\label{sec_homWE}
\subsubsection{Vector potential}\label{subsec_homWEA}

Generally, in order to derive electromagnetic wave equations in media, one has to start from the fundamental, Lorentz-covariant equation of motion \eqref{fund_wave_eq} for the total four-potential, split the total current into induced and external contributions,
\begin{equation}
 j^\mu \equiv j^\mu_{\rm tot} = j^\mu_{\rm ext} + j^\mu_{\rm ind} \,,
\end{equation}
then set the external contributions to zero,
\begin{equation}
 j^\mu_{\rm ext} := 0 \,,
\end{equation}
and eliminate the induced four-current density by means of the proper fundamental response tensor as
\begin{equation}
j^\mu_{\rm ind}=\widetilde\chi\indices{^\mu_\nu} \, A^\nu_{\rm tot} \,.
\end{equation}
This logic has been discussed paradigmatically by the authors of this article in Ref.~\cite{Refr} and will therefore be taken for granted for the purposes of this article.
From the above procedure, we obtain the fundamental wave equation for the total four-potential $A^\mu \equiv A^\mu_{\rm tot}$ in the form
\begin{equation} \label{eq_FundWaveMedia_realspace_2}
 (\eta\indices{^\mu_\nu}\Box+\partial^\mu\partial_\nu - \mu_0 \h \widetilde \chi\indices{^\mu_\nu}\h) \h A^\nu = 0 \,.
\end{equation}
In Fourier space, this can be written equivalently as \cite[Eq.~(4.7)]{Refr}
\begin{equation}\label{eq_FundWaveMedia}
\bigg(\bigg({-\frac{\omega^2}{c^2}}+|\vec k|^2 \bigg) \h \eta\indices{^\mu_\nu}-k^\mu k_\nu-\mu_0\,\widetilde \chi\indices{^\mu_\nu}(\vec k,\omega)\bigg) \h A^\nu(\vec k,\omega)=0\,.
\end{equation}
In the temporal gauge $A^0 \equiv \varphi/c = 0$, this further reduces to \cite[\S\,4.1.2]{Refr}
\begin{equation} \label{wave_temporal}
 \bigg({-\frac{\omega^2}{c^2}} + |\vec k|^2 -\mu_0 \, \bigg( \tsr 1 - \frac{c^2 |\vec k|^2}{\omega^2} \h \tsr P_{\rm L}(\vec k) \bigg) \tsr{\widetilde \chi}(\vec k, \omega)\h \bigg) \h \vec A(\vec k, \omega) = 0 \,,
\end{equation}
which is the desired wave equation for the vector potential in materials. In particular, for the case of a vanishing current response tensor, we recover from it the free wave equation. We remark that an alternative derivation of Eq.~\eqref{wave_temporal} can be given by starting from the equation of motion for the vector potential, Eq.~\eqref{wave_A_equiv}, which is equivalent to
\begin{equation}
\bigg({-\frac{\omega^2}{c^2}} + |\vec k|^2\bigg) \h \vec A(\vec k,\omega)=\mu_0 \,\bigg(\tsr 1-\frac{c^2|\vec k|^2}{\omega^2} \h \tsr P\L(\vec k)\bigg) \, \vec j(\vec k,\omega)\,,
\end{equation}
and by eliminating the spatial current density directly via
\begin{equation}
\vec j(\vec k,\omega)=\tsr{\widetilde\chi}(\vec k,\omega) \h \vec A(\vec k,\omega)
\end{equation}
in terms of the proper current response tensor.

To further simplify the wave equation, we factor out the free wave operator~and thus rewrite Eq.~\eqref{wave_temporal} as
\begin{equation}
\mathbbmsl D_0^{-1}(\vec k,\omega) \, \bigg(\tsr 1-\mathbbmsl D_0(\vec k,\omega) \, \bigg(\tsr 1-\frac{c^2|\vec k|^2}{\omega^2} \h \tsr P\L(\vec k)\bigg) \h \tsr{\widetilde\chi}(\vec k,\omega)\bigg) \h \vec A(\vec k,\omega)=0\,.
\end{equation}
On account of the expression \eqref{eq_GFtempGauge} for the free Cartesian Green function, this can be transformed into
\begin{equation} \label{eq_epsA=0_alt}
\mathbbmsl D_0^{-1}(\vec k,\omega) \, \Big( \h \tsr 1-\tsr D_0(\vec k,\omega) \h \tsr{\widetilde\chi}(\vec k,\omega)\Big) \h \vec A(\vec k,\omega)=0\,.
\end{equation}
Using the representation \eqref{eq_eps} of the dielectric tensor, we finally obtain
\begin{equation}
\mathbbmsl D_0^{-1}(\vec k,\omega) \h \tsr{\varepsilon}_{\rm r}(\vec k,\omega) \h \vec A(\vec k,\omega)=0\,. \label{eq_epsA=0}
\end{equation}
From this result, we will derive in the following subsection the fundamental wave equation for the electric field in materials.

\subsubsection{Electric field}

The electric field is related to the vector potential in the temporal gauge by Eq.~\eqref{eq_EA}. Consequently, by simply
multiplying both sides of Eq.~\eqref{eq_epsA=0} with $\i\omega$\h, the wave equation for the electric field in materials turns out to be
\begin{equation}
\mathbbmsl D_0^{-1}(\vec k,\omega) \h \tsr\varepsilon_{\rm r}(\vec k,\omega) \h \vec E(\vec k,\omega)=0\,. \label{eq_BoxEpsE=0}
\end{equation}
If we set $\varepsilon_{\rm r}=1$
in this equation---which corresponds to the vacuum case---then we recover the free wave equation 
\begin{equation}
\bigg({-\frac{\omega^2}{c^2}} + |\vec k|^2 \bigg) \h \vec E(\vec k,\omega) = 0\,. 
\end{equation}
By contrast, if we exclude the free dispersion relation $\omega=c|\vec k|$, then the inverse free Green function in Eq.~\eqref{eq_BoxEpsE=0} is non-zero and can therefore be canceled, which results in the simple equation
\begin{equation}
 \tsr\varepsilon_{\rm r}(\vec k,\omega) \h \vec E(\vec k,\omega)=0\,. \label{eq_epsE=0}
\end{equation}
This is the fundamental, homogeneous electromagnetic wave equation in materials, 
stating that the electric field in materials is restricted to the null space of the dielectric tensor. Apparently, it was first discovered 
by O.\,V.~Dolgov and E.\,G.~Maksimov \cite[Eq.~(2.34) and comments below]{Dolgov}. Later, a paradigmatic discussion has been given by the authors of this article in Ref.~\cite{Refr}. 

We stress that apart from the linear behaviour of the material, Eq.~\eqref{eq_epsE=0} only assumes the possibility of treating the material as homogeneous. Otherwise, it includes
all the effects of  anisotropy, relativistic retardation, and magneto\-{}electric cross coupling. In particular, being valid in {\itshape all} materials, it applies to both metals and insulators.

Finally, we remark that Eq.~\eqref{eq_epsE=0} can be derived independently by starting from the fundamental equation of motion for the electric field with sources, Eq.~\eqref{eq_FundWaveSourFourier}. For this purpose, one eliminates the current density via Ohm's law in terms of the proper conductivity tensor,
\begin{equation}
\vec j(\vec k,\omega)=\tsr{\widetilde\sigma}(\vec k,\omega) \h \vec E(\vec k,\omega)\,, \label{eq_OhmLaw}
\end{equation}
and similarly the charge density by means of the continuity equation,
\begin{equation}
\rho(\vec k,\omega)=\frac{1}{\omega}\h \vec k^{\rm T} \h \tsr{\widetilde\sigma}(\vec k,\omega)\h \vec E(\vec k,\omega)\,.
\end{equation}
Using the relations derived in \S\,\ref{subsec_DiTens}, this leads again to Eq.~\eqref{eq_epsE=0}.

\subsubsection{Current density}

In the preceding subsections, we have rederived the wave equations for the vector potential and the electric field in materials, which describe proper
oscillations of the medium (i.e., oscillations in the absence of external perturbations). These proper oscillations are generated by the charge and current fluctuation of the medium itself and therefore
completely determined by these. In this subsection, we will also derive the wave equation for these charge and current fluctuations.

We start again from the fundamental, Lorentz-covariant wave equation for the four-potential, Eq.~\eqref{eq_FundWaveMedia_realspace_2}, and eliminate the four-potential in terms of the four-current by means of the relation \cite[\S\,3.3]{ED1} \smallskip
\begin{equation} \label{gfeq}
 A^\mu_{\rm tot} = (D_0)\indices{^\mu_\nu} \, j^{\nu}_{\rm tot} + \partial^\mu \mh f\,, \smallskip
\end{equation}
where $(D_0)\indices{^\mu_\nu}$ denotes the free electromagnetic Green function, and $\partial^\mu \mh f$ is any pure gauge. Then, the pure gauges drop out of the equation, since they always solve the homogeneous wave equation in materials,
\begin{equation}
  (\eta\indices{^\mu_\nu}\Box+\partial^\mu\partial_\nu - \mu_0 \h \widetilde \chi\indices{^\mu_\nu}\h) \, \partial^\nu \! f = 0 \,.
\end{equation}
On the other hand, by definition of the tensorial Green function, the identity
\begin{equation}
 (\eta\indices{^\mu_\nu}\Box+\partial^\mu\partial_\nu) \h (D_0)\indices{^\nu_\lambda} \, j^\lambda = \mu_0 \h j^\mu
\end{equation}
holds for any physical (i.e., Minkowski-transverse) four-current $j^\mu$ \cite[\S\,3.3]{ED1}. Thus, we obtain the wave equation for the total four-current, $j^\mu \equiv j^\mu_{\rm tot}$\hh, as
\begin{equation} \label{wave_curr_1}
 (\eta\indices{^\mu_\nu} - \widetilde \chi\indices{^\mu_\lambda} \h (D_0)\indices{^\lambda_\nu}\h ) \h\hh j^\nu = 0 \,.
\end{equation}
We now put in the general form of the free electromagnetic Green function derived in \cite[Eq.~(3.41)]{ED1}, i.e.,
\begin{equation}
 (D_0)\indices{^\mu_\nu}(k) = \mathbbmsl D_0(k) \left( \eta\indices{^\mu_\nu} + \frac{c k^\mu}{\omega} \, f_\nu(k) + g^\mu(k) \, \frac{c k_\nu}{\omega} + \frac{c k^\mu}{\omega} \, h(k) \, \frac{c k_\nu}{\omega} \right),
\end{equation}
where $f_\nu$, $g^\mu$ and $h$ are arbitrary complex functions up to the constraints
\begin{equation}
 f_\nu(k) \, k^\nu = k_\mu \, g^\mu(k) = 0 \,.
\end{equation}
Further using the constraints \eqref{eq_constraints}
on the fundamental response tensor as well as the continuity equation \eqref{cont_eq_covar},
we see that in Eq.~\eqref{wave_curr_1} all terms 
containing the arbitrary functions $f_\nu$, $g^\mu$ and $h$ drop out, thus leading to
\begin{equation} \label{eq_FundWaveMedia_curr}
 (\eta\indices{^\mu_\nu} - \widetilde \chi\indices{^\mu_\nu} \h \mathbbmsl D_0 \hh ) \h\hh  j^\nu = 0 \,.
\end{equation}
Multiplying through with $\mathbbmsl D_0^{-1}$, we arrive at
\begin{equation}
 \bigg(\eta\indices{^\mu_\nu}\h \bigg({-\frac{\omega^2}{c^2}+|\vec k|^2}\bigg) - \mu_0 \h \widetilde\chi\indices{^\mu_\nu}(\vec k,\omega)\hh \bigg) \,  j^\nu(\vec k,\omega) = 0 \,.
\end{equation}
This is the general, microscopic, manifestly Lorentz-covariant wave equation for the electromagnetic four-current in materials.

Similarly as in the case of the four-potential, we can rewrite the wave equation \eqref{eq_FundWaveMedia_curr} for $j^\mu = (c\rho, \h \vec j)^{\rm T}$ 
in terms of the charge density, the spatial current density, and the spatial part of the proper response tensor. 
Using Eq.~\eqref{generalform1}, we thereby obtain
\begin{align}
 \bigg( {- \frac{c\vec k^{\rm T}}{\omega} \, \tsr{\widetilde \chi} \, \mathbbmsl D_0 } \bigg) \h\hh \vec j + \bigg( 1 + \frac{c\vec k^{\rm T}}{\omega} \, \tsr{\widetilde \chi} \,\h \frac{c\vec k}{\omega} \, \mathbbmsl D_0 \bigg) \h c\hh \rho & = 0 \,, \\[3pt]
 \bigg(\tsr 1 - \tsr{\widetilde \chi} \, \mathbbmsl D_0\bigg) \h\hh \vec j + \bigg( \h \tsr {\widetilde \chi} \,\h \frac{c\vec k}{\omega} \, \mathbbmsl D_0 \bigg) \h c\hh\rho & = 0 \,. \label{eq_wave_current}
\end{align}
These equations are again not independent of each other: multiplying the second one from left with $c\vec k^{\rm T}/\omega$ and using the continuity equation 
yields \linebreak the first equation, which can hence be discarded.
To simplify matters further, we eliminate the charge density $\rho$ completely from Eq.~\eqref{eq_wave_current} by means of the continuity equation (analogously as in \S\,\ref{sec_cart_form}). Thereby, we obtain
\begin{equation} \label{zwischen_1}
 \left( \tsr 1 - \tsr{\widetilde \chi}(\vec k, \omega) \h \mathbbmsl D_0(\vec k, \omega) \left( \tsr 1 - \frac{c^2|\vec k|^2}{\omega^2} \, \tsr P_{\rm L}(\vec k) \right) \right) \vec j(\vec k, \omega) = 0 \,.
\end{equation}
With the explicit form \eqref{eq_GFtempGauge} of the Cartesian Green function, we can write this compactly as 
\begin{equation} \label{eq_hom_curr}
 \Big( \h \tsr 1 - \tsr{\widetilde \chi}(\vec k, \omega) \h \tsr D_0(\vec k, \omega) \Big) \h \vec j(\vec k, \omega) = 0 \,.
\end{equation}
Finally, by multiplying Eq.~\eqref{zwischen_1} through with $\mathbbmsl D_0^{-1}$, we arrive at
\begin{equation} \label{eq_WEcurrent}
 \left(-\frac{\omega^2}{c^2}+|\vec k|^2 - \mu_0 \h \tsr{\widetilde \chi}(\vec k, \omega) \left( \tsr 1 - \frac{c^2|\vec k|^2}{\omega^2} \, \tsr P_{\rm L}(\vec k) \right) \right) \vec j(\vec k, \omega) = 0 \,.
\end{equation}
This is the fundamental wave equation for the electromagnetic current in materials. It is also Lorentz covariant because it has been derived from the manifestly Lorentz-covariant equation \eqref{eq_FundWaveMedia_curr}. We further note that the homogeneous wave equation for the spatial current, Eq.~\eqref{eq_hom_curr}, differs from the  homogeneous  wave equation for the vector potential, Eq.~\eqref{eq_epsA=0_alt}, only in the order of the Cartesian Green function $D_0$ and the proper current response tensor $\widetilde\chi$ (and in the additional factor $\mathbbmsl D_0$ which, however, can be canceled by excluding the free dispersion relation).

\subsection{Inhomogeneous wave equations} \label{sec_inhomWE}

In this subsection, we derive an even more general class of electromagnetic wave equations in media.
In fact, in the conventional approach explained in the preceding subsection, one starts from the fundamental
inhomogeneous wave equations and eliminates the source terms by means of proper response functions.
In actual fact though, this is only possible if the external sources vanish, as would be the case, for example, for externally applied free light waves.
More generally, one has to allow for the presence of external sources (describing e.g.~impurities or charged projectiles moving through the medium), 
and these can as a matter of principle not be eliminated
by linear response theory. Correspondingly, we derive in this subsection {\it inhomogeneous} wave equations for the {\it total} electromagnetic fields with {\it external} sources.
By setting the external sources to zero in the final results, one then recovers the corresponding homogeneous wave equations derived above.

\subsubsection{Vector potential}

As we have shown already, any vector potential can be expressed in terms of its own sources by means of the {\itshape free} Green function. The  corresponding relation \eqref{eq_EoMASol} can be formally inverted as
\begin{equation}
(\tsr D_0)^{-1}(\vec k,\omega)\vec A(\vec k,\omega)=\vec j(\vec k,\omega)\,,
\end{equation}
and this in turn is equivalent to the fundamental wave equation \eqref{eq_EoMA1} for the vector potential (see Eq.~\eqref{eq_InvCartGF}).
On the other hand, in the context of electrodynamics in media, the {\itshape total} vector potential can also be expressed in terms of the {\itshape external} current by means of the {\itshape full} Green function. The corresponding relation \eqref{eq_ADjext} can again be formally inverted and thereby yields the formal wave equation
\begin{equation}
\tsr D{}^{-1}(\vec k,\omega) \h \vec A(\vec k,\omega)=\vec j\ext(\vec k,\omega)\,. \label{eq_inhomoWEAj}
\end{equation}
This is the {\itshape inhomogeneous} wave equation for the vector potential in materials. Using the relation between the 
full Cartesian Green function and the dielectric tensor, Eq.~\eqref{eq_DvsEps}, we see that it is equivalent to
\begin{equation}
 (\tsr D_0)^{-1}(\vec k,\omega) \h \tsr\varepsilon_{\rm r}(\vec k,\omega) \h \vec A(\vec k,\omega) = \vec j_{\rm ext}(\vec k, \omega) \,. \label{eq_inhomoWEAj_equiv}
\end{equation}
In particular, in the absence of external sources, we recover from this the {\itshape homogeneous} wave equation for the vector potential in media, Eq.~\eqref{eq_epsA=0}.

\subsubsection{Electric field}

The inhomogeneous wave equation for the electric field in media can be derived from Eq.~\eqref{eq_inhomoWEAj_equiv} by using again the relation \eqref{eq_EA} between the electric field and the vector potential in the temporal gauge. It is given by
\begin{equation}\label{eq_inhomoWE}
\tsr\varepsilon_{\rm r}(\vec k,\omega) \h \vec E(\vec k,\omega)=\i\omega \h \tsr D_0(\vec k,\omega) \h\hh \vec j\ext(\vec k,\omega)\,,
\end{equation}
which obviously reduces to Eq.~\eqref{eq_epsE=0} for vanishing external current. An alternative derivation of this inhomogeneous wave equation can be given by starting from the relation \eqref{el_sol}  for the external quantities and using the explicit expression \eqref{el_sol_exp} for the electric solution generator:
\begin{equation} \label{zwischen_4}
 \vec E\ext \h = \h \frac{1}{\i\omega\varepsilon_0} \, \tsr{\mathbbmsl E} \h\hh \vec j\ext \h = \h \i\omega \hh \tsr D_0 \, \vec j_{\rm ext} \,.
\end{equation}
Further eliminating the external electric field as
\begin{equation}
\vec E\ext \h = \h \tsr\varepsilon_{\rm r}\h\vec E\tot \h \equiv \h \tsr\varepsilon_{\rm r}\h\vec E\,,
\end{equation}
we thus recover the result \eqref{eq_inhomoWE}.
Finally, we rewrite Eq.~\eqref{eq_inhomoWE} in terms of the proper current response tensor (see Eq.~\eqref{eq_eps}) as
\begin{equation} \label{zwischen_2b}
\Big( \h \tsr 1-\tsr D_0 \h \tsr{\widetilde\chi}\h \Big) \h \vec E= \i\omega \hh \tsr D_0 \, \vec j\ext \,.
\end{equation}
Using Eq.~\eqref{eq_GFtempGauge} for the free Cartesian Green function and multiplying through with $\mu_0 \hh \mathbbmsl D_0^{-1}$, we obtain the explicit wave equation
\begin{equation} \label{compare_this}
\begin{aligned}
 \bigg( {-\frac{\omega^2}{c^2}} + |\vec k|^2 - \mu_0 \, \bigg( \tsr 1 - \frac{c^2|\vec k|^2}{\omega^2} \, \tsr P_{\rm L}(\vec k) \bigg) \h \tsr{\widetilde \chi}(\vec k, \omega) \bigg) \h \vec E(\vec k, \omega) = & \\[3pt]
 \i\omega \h \mu_0\, \bigg( \tsr 1 - \frac{c^2|\vec k|^2}{\omega^2} \, \tsr P_{\rm L}(\vec k) \bigg) \, \vec j\ext(\vec k, \omega) \,, &
\end{aligned}
\end{equation}

\pagebreak \noindent
which will be useful later for the comparison with the inhomogeneous wave equation for the current density.

\subsubsection{Current density}

Finally, in order to derive the inhomogeneous wave equation for the current density, we represent the total electric field in Eq.~\eqref{eq_inhomoWE} in terms of its generating current analogously as in Eq.~\eqref{zwischen_4}, hence,
\begin{equation}
 \tsr \varepsilon_{\rm r} \h \tsr D_0 \, \vec j = \tsr D_0 \, \vec j\ext \,.
\end{equation}
With the connection \eqref{eq_eps} between the dielectric tensor and the proper current response tensor, this reverts to
\begin{equation}
 \Big(\h \tsr 1 - \tsr D_0 \h \tsr{\widetilde \chi} \h \Big) \hh \tsr D_0 \, \vec j = \tsr D_0 \, \vec j\ext \,.
\end{equation}
By inverting the free Green function, we further obtain
\begin{equation} \label{zwischen_2a}
 \Big( \h \tsr 1 - \tsr{\widetilde \chi} \h \tsr D_0 \h \Big) \h\hh \vec j = \vec j\ext \,,
\end{equation}
and this obviously reduces to the homogenous equation \eqref{eq_hom_curr} for vanishing external current. Finally, by multiplying through with $\mu_0 \hh \mathbbmsl D_0^{-1}$ and using Eq. \eqref{eq_GFtempGauge}, we arrive at
\begin{equation}
\begin{aligned} \label{zwischen_2}
\left(-\frac{\omega^2}{c^2}+|\vec k|^2 -  \mu_0 \h \tsr{\widetilde\chi}(\vec k, \omega) 
\left(\tsr 1 - \frac{c^2|\vec k|^2}{\omega^2} \, \tsr P_{\rm L}(\vec k) \right) \right) \vec j(\vec k, \omega) = & \\[3pt]
\left( -\frac{\omega^2}{c^2} + |\vec k|^2 \right) \vec j\ext(\vec k,\omega) \,. &
\end{aligned}
\end{equation}
This is the general, inhomogeneous wave equation for the current density in materials. It is precisely the inhomogeneous version of Eq.~\eqref{eq_WEcurrent}, and it may be compared to the corresponding inhomogeneous wave equation for the electric field  given by Eq.~\eqref{compare_this}.

\section{Isotropic limit} \label{sec_iso}

\subsection{Longitudinal and transverse response functions} \label{longtranseps}

In this final section, we shortly reassemble our results under the assumption of the {\itshape isotropic limit,} which is important for most practical applications. {\it Per definitionem}, in the isotropic limit the direct current response tensor assumes the form
\begin{equation}
\tsr{\chi}(\vec k,\omega)=\chi\L(\vec k,\omega)\tsr P\L(\vec k) + \chi\T(\vec k,\omega)\tsr P\T(\vec k)\,,
\end{equation}
where $\chi\L$ and $\chi\T$ respectively denote the {\itshape longitudinal} and {\itshape transverse} current response functions (see \S\,\ref{sec_proj}). On the other hand, from the general form \eqref{generalform1} of the fundamental response tensor, we read off directly the general relation between the {\itshape density response function} $\upchi$ and the current response tensor,
\begin{equation} 
 \upchi(\vec k, \omega) \equiv \frac{1}{c^2} \h \chi\indices{^0_0}(\vec k,\omega) = -\frac{1}{\omega^2} \, \vec k^{\rm T} \, \tsr\chi(\vec k, \omega) \h \vec k \,.
\end{equation}
By combining these equations, we obtain the connection between the density response function and the longitudinal current response function,
\begin{equation} \label{eq_DensRespLongCurr}
 \upchi(\vec k, \omega) = -\frac{|\vec k|^2}{\omega^2} \h \chi\L(\vec k,\omega) \,.
\end{equation}
Furthermore, the Cartesian Dyson equation \eqref{Cart_dyson_1} together with the representation \eqref{freeCart_decom}--\eqref{freeCart_trans} of the free Cartesian Green function imply that in the isotropic limit, the {\itshape proper} current response tensor is of the form
\begin{equation} \label{iso_proper}
\tsr{\widetilde \chi}(\vec k,\omega)=\widetilde\chi\L(\vec k,\omega)\tsr P\L(\vec k) + \widetilde\chi\T(\vec k,\omega)\tsr P\T(\vec k)\,,
\end{equation}
and the analogous relation \eqref{eq_DensRespLongCurr} also holds between the {\itshape proper} density response  function $\widetilde{\upchi}$ and the {\itshape proper} longitudinal current response function $\widetilde{\chi}_{\rm L}$. In particular, the direct and the proper density response functions satisfy the Dyson-type equation
\begin{equation}
\upchi(\vec k, \omega) = \widetilde \upchi(\vec k, \omega) + \widetilde \upchi(\vec k, \omega) \, v(\vec k) \, \upchi(\vec k, \omega) \,,
\end{equation}
where $v(\vec k) = 1/(\varepsilon_0 \hh |\vec k|^2)$ is the Coulomb kernel in Fourier space. This equation is well-known 
in ab initio electronic structure physics \cite[Eq.~(5.17)]{Giuliani} and is a special case of the more general relation \eqref{Cart_dyson_1}.

Next, Eq.~\eqref{eq_eps} implies that the dielectric tensor is also isotropic, 
\begin{equation}
\tsr\varepsilon_{\rm r}(\vec k,\omega)=\varepsilon_{\rm r, \hh L}(\vec k,\omega)\tsr P\L(\vec k) + \varepsilon_{\rm r, \hh T}(\vec k,\omega)\tsr P\T(\vec k)\,,
\end{equation}
where the {\itshape longitudinal} and the {\itshape transverse dielectric function} are respectively related to the proper density response function and the proper transverse current response function as
\begin{align}
\varepsilon_{\rm r, \hh L}(\vec k,\omega)&= 1-v(\vec k) \h \widetilde\upchi(\vec k,\omega)\,, \label{diel_L} \\[5pt]
\varepsilon_{\rm r, \hh T}(\vec k,\omega)&=1-\mathbbmsl D_0(\vec k,\omega) \h \widetilde\chi\T(\vec k,\omega)\,. \label{diel_T}
\end{align}
Similarly, from Eq.~\eqref{eq_epsinv} one obtains the connection between the inverse dielectric functions and the corresponding direct response functions,
\begin{align}
\varepsilon^{-1}_{\rm r, \hh L}(\vec k,\omega)&=1+v(\vec k)\h \upchi(\vec k,\omega)\,,\label{eq_longDiFct}\\[5pt]
\varepsilon^{-1}_{\rm r, \hh T}(\vec k,\omega)&=1+\mathbbmsl D_0(\vec k,\omega) \h \chi\T(\vec k,\omega)\,. \label{eq_transDiFct}
\end{align}
By combining these equations, we further obtain the relations
\begin{align}
 \widetilde \upchi(\vec k, \omega) & = \upchi(\vec k, \omega) \, \varepsilon_{\rm r, \hh L}(\vec k, \omega) \,, \label{some_rel} \\[5pt]
\widetilde \chi\T(\vec k, \omega) & = \chi\T(\vec k, \omega) \, \varepsilon_{\rm r, \hh T}(\vec k, \omega) \,,
\end{align}
which also follow directly from the more general tensorial relation \eqref{tildechichieps}.

In this context, we remark that the ``generalized dielectric constant'' $\varepsilon_{\rm r}$, which is often defined as (see e.g.~\cite[Eq.~(E.14)]{Martin})
\begin{equation}
 \varepsilon_{\rm r}(\vec k, \omega) \stackrel{?}{=} \frac{\delta\varphi\ext(\vec k, \omega)}{\delta\varphi\tot(\vec k, \omega)}\,, \label{eq_defdielectricfct}
\end{equation}
generally has to be identified with the {\itshape longitudinal} dielectric function $\varepsilon_{\rm r, \hh L}$ in the isotropic limit. In fact, the problematic feature of the above definition \eqref{eq_defdielectricfct} is that this is in general not equivalent to
\begin{equation}
 \varphi\ext(\vec k,\omega)=\varepsilon_{\rm r}(\vec k,\omega) \h \varphi\tot(\vec k,\omega)\,, \label{eq_defdielectricfctexp}
\end{equation}
unless the vector potential vanishes identically, $\vec A \equiv 0$. In the most general case, the dielectric {\itshape tensor} has of course to be introduced via the linear expansion of the external
in terms of the total electric field as in Eq.~\eqref{eq_DefDiTens}.
Although the dielectric function $\varepsilon_{\rm r}$ is therefore actually a {\itshape longitudinal} response function, it is used in electronic structure theory
for the description of {\itshape optical} properties---which are induced by {\itshape transverse} electromagnetic fields. An explanation for this astonishing fact is given in Ref.~\cite[\S\,4.4]{Refr}.

\pagebreak
Finally, from the Cartesian Dyson equation in the form \eqref{eq_CartDyson_2} and from Eqs.~\eqref{freeCart_long}--\eqref{freeCart_trans} it follows that the full Cartesian Green function is given in the isotropic limit by
\begin{equation}
\tsr D(\vec k,\omega)=D\L(\vec k,\omega)\tsr P\L(\vec k)+
D\T(\vec k,\omega)\tsr P\T(\vec k)\,,
\end{equation}
with the scalar coefficient functions
\begin{align}
D\L(\vec k,\omega)&=\frac{\mu_0}{-\omega^2/c^2-\mu_0\h\widetilde\chi\L(\vec k,\omega)}\,,\\[6pt]
D\T(\vec k,\omega)&=\frac{\mu_0}{-\omega^2/c^2+|\vec k|^2-\mu_0\h\widetilde\chi\T(\vec k,\omega)}\,. \label{bis_hier}
\end{align}
In the following, we will use these equations to formulate the corresponding wave equations in the isotropic limit.

\subsection{Decoupled wave equations}

In the isotropic limit, where each response tensor is determined by only one longitudinal and one transverse response function, 
the wave equations derived so far also decouple into separate equations for the longitudinal and transverse components of the corresponding field quantities. 
For deriving these decoupled wave equations, we start from the general inhomogeneous wave equation for the current density, Eq.~\eqref{zwischen_2a}, 
or for the electric field, Eq.~\eqref{zwischen_2b}. These two equations differ only in the order of the proper current response tensor and the free Cartesian Green function  in the respective wave operators, as well as in the additional operator $(\i\omega D_0)$ on the right-hand side of Eq.~\eqref{zwischen_2b}.
However, the order of the two tensors $\widetilde \chi$ and $D_0$  does not play any r\^{o}le in the isotropic limit, where formally,
\begin{equation}
 \tsr{\widetilde\chi} \h \tsr D_0 \h = \h \widetilde \chi\L \h D_{0, \hh \mathrm L} \, \tsr P\L + \widetilde \chi\T \h D_{0, \hh \mathrm T} \, \tsr P\T \h = \h \tsr D_0 \h \tsr{\widetilde\chi} \,.
\end{equation}
Correspondingly,  we first obtain from Eq.~\eqref{zwischen_2a} the decoupled wave equations for the longitudinal and transverse current densities,
\begin{align}
 \big(1 - \widetilde \chi\L \h D_{0, \hh \rm L} \big) \h \vec j\L & = \vec j_{\rm ext, \hh L} \,,\\[5pt]
 \big(1 - \widetilde \chi\T \h D_{0, \hh \rm T}\big) \h \vec j\T & = \vec j_{\rm ext, \hh T} \,.
\end{align}
By Eqs.~\eqref{freeCart_long}--\eqref{freeCart_trans}, these are equivalent to
\begin{align}
 \left( -\frac{\omega^2}{c^2} - \mu_0 \h \widetilde \chi\L(\vec k, \omega) \right) \vec j_{\rm L}(\vec k, \omega) & = -\frac{\omega^2}{c^2} \h \vec j_{\rm ext, \hh L}(\vec k, \omega) \,,\\[3pt]
 \left( -\frac{\omega^2}{c^2} + |\vec k|^2 - \mu_0 \h \widetilde \chi\T(\vec k, \omega) \right) \vec j_{\rm T}(\vec k, \omega) & = \left( -\frac{\omega^2}{c^2} + |\vec k|^2 \right) \vec j_{\rm ext, \hh T}(\vec k, \omega) \,.
\end{align}
In terms of the longitudinal and transverse dielectric functions \eqref{diel_L}--\eqref{diel_T}, these wave equations take the particularly simple form
\begin{align}
 \varepsilon_{\rm r, \hh L}(\vec k, \omega) \, \vec j\L(\vec k, \omega) & = \vec j_{\rm ext, \hh L}(\vec k, \omega) \,, \label{simple_l} \\[5pt]
 \varepsilon_{\rm r, \hh T}(\vec k, \omega) \, \vec j\T(\vec k, \omega) & = \vec j_{\rm ext, \hh T}(\vec k, \omega) \,. \label{simple_t}
\end{align}
Furthermore, with the relation between the longitudinal current density and the charge density, \smallskip
\begin{equation}
\vec j\L(\vec k,\omega)=\frac{\omega \hh \rho(\vec k,\omega)}{|\vec k|^2} \h \vec k\,, \label{eq_continuityFourier} \smallskip
\end{equation}
we obtain the inhomogeneous wave equation for the charge density,
\begin{equation}
 \varepsilon_{\rm r, \hh L}(\vec k, \omega) \h \rho(\vec k, \omega) = \rho_{\rm ext, \hh L}(\vec k, \omega) \,.
\end{equation}
In the absence of external sources, this reduces to the well-known {\itshape plasmon equation} (see e.g.~\cite[Eq.~(4.91)]{MartinRothen})
\begin{equation}
 \varepsilon_{\rm r, \hh L}(\vec k, \omega) \, \rho(\vec k, \omega) = 0 \,.
\end{equation}
Next, for the electric field we obtain from Eq.~\eqref{zwischen_2b} in the isotropic limit the decoupled wave equations
\begin{align}
 \big(1 - D_{0, \hh \rm L} \h \widetilde \chi\L \big) \h \vec E\L & = \i\omega \h \bigg({-\frac{1}{\varepsilon_0 \h \omega^2}} \bigg) \h \vec j_{\rm ext, \hh L} \,,\\[3pt]
 \big(1 - D_{0, \hh \rm T} \h \widetilde \chi\T \big) \h \vec E\T & = \i\omega \h \mathbbmsl D_0 \, \vec j_{\rm ext, \hh T} \,.
\end{align}
These are in turn equivalent to
\begin{align}
 \left( -\frac{\omega^2}{c^2} - \mu_0 \h \widetilde \chi\L(\vec k, \omega) \right) \mh \vec E_{\rm L}(\vec k, \omega) & = \i\omega \h\hh \mu_0 \h\hh \vec j_{\rm ext, \hh L}(\vec k, \omega) \,,\\[3pt]
 \left( -\frac{\omega^2}{c^2} + |\vec k|^2 - \mu_0 \h \widetilde \chi\T(\vec k, \omega) \right) \mh \vec E_{\rm T}(\vec k, \omega) & = \i\omega \h\hh \mu_0 \h\hh  \vec j_{\rm ext, \hh T}(\vec k, \omega) \,,
\end{align}
or in terms of the longitudinal and transverse dielectric functions,
\begin{align}
 -\frac{\omega^2}{c^2} \, \varepsilon_{\rm r, \hh L}(\vec k,\omega) \h \vec E\L(\vec k,\omega)				&= \i\omega \h\hh \mu_0 \h\hh \vec j_{\rm ext, \hh L}(\vec k, \omega) \,, \\[3pt]
 \left(-\frac{\omega^2}{c^2}+|\vec k|^2\right) \mh \varepsilon_{\rm r, \hh T}(\vec k,\omega) \h \vec E\T(\vec k,\omega)	&= \i\omega \h\hh \mu_0 \h\hh  \vec j_{\rm ext, \hh T}(\vec k, \omega) \,.
\end{align}
For vanishing external sources, we thus retrieve the simple equations
\begin{align}
 \varepsilon_{\rm r, \hh L}(\vec k, \omega) \h \vec E\L(\vec k,\omega) & = 0 \,,\\[5pt]
\varepsilon_{\rm r, \hh T}(\vec k, \omega) \h \vec E\T(\vec k,\omega) & = 0 \,,
\end{align}
which respectively describe the propagation of longitudinal plasmons and transverse light waves (see Ref.~\cite[\S\,4.3]{Refr}). 
In fact, in the absence of external sources these wave equations are exactly analogous to the corresponding homogeneous wave equations for the current density (Eqs.~\eqref{simple_l}--\eqref{simple_t} with $\vec j_{\rm ext} = 0$). This is consistent with the fact that
in the isotropic limit, the various field quantities are simply interrelated by
\begin{align}
\vec j\L(\vec k,\omega)&=\widetilde\chi\L(\vec k,\omega)\vec A\L(\vec k,\omega)=\frac{1}{\i\omega} \, \widetilde\chi\L(\vec k,\omega)\vec E\L(\vec k,\omega)\,,\\[5pt]
\vec j\T(\vec k,\omega)&=\widetilde\chi\T(\vec k,\omega)\vec A\T(\vec k,\omega)=\frac{1}{\i\omega} \, \widetilde\chi\T(\vec k,\omega)\vec E\T(\vec k,\omega)\,.
\end{align}
Finally, the decoupled wave equations for the longitudinal and transverse vector potential in the temporal gauge can be obtained from the respective equations for the electric field by means of the relation $\vec E = \i\omega \vec A$.

\section{Conclusion}

We have presented a systematic and comprehensive investigation of linear electromagnetic wave equations in materials. For this purpose, we have first given a short introduction to microscopic electrodynamics in media, and then investigated the relations of the microscopic dielectric tensor to other response functions.
Our main results in this context are: the relation \eqref{eq_eps} between the dielectric tensor and the proper current response tensor, the  relation \eqref{eq_epsinv} between the inverse dielectric tensor and the direct current response tensor, and the relation \eqref{eq_DvsEps} between the dielectric tensor and the full electromagnetic Green function. In particular, the tensorial equations \eqref{eq_eps} and \eqref{eq_epsinv} generalize the respective scalar equations \eqref{diel_L} and \eqref{eq_longDiFct}, which are already well-known in {\itshape ab initio} electronic structure physics.

Subsequently, we have given a review of homogeneous microscopic wave equations in materials (\S\,\ref{sec_homWE})
and derived their inhomogeneous counterparts (\S\,\ref{sec_inhomWE}), which describe electromagnetic oscillations in materials driven by external currents. Concretely, 
the most general wave equations for the electromagnetic vector potential, the electric field and the current density are given by 
Eqs.~\eqref{eq_inhomoWEAj_equiv}, \eqref{compare_this} and \eqref{zwischen_2}, respectively. Finally, we have investigated these formulae in the isotropic limit in Sct.~\ref{sec_iso}.

As this article quite generally derives linear electromagnetic wave equations in materials from first principles, it paves the way for the application
of {\itshape ab initio} results to the problem of wave and pulse propagation in media.

\section*{Acknowledgments}
This research was supported by the DFG grant HO 2422/12-1. R.\,S. thanks the Institute for Theoretical Physics at TU Bergakademie Freiberg for its hospitality.

\bigskip
\bigskip
\noindent
{\bfseries References}
\bibliographystyle{model1-num-names}
\bibliography{/net/home/lxtsfs1/tpc/schober/Ronald/masterbib}

\end{document}